\documentclass{he_symp}
\usepackage{psfig,graphicx,epsfig}
\usepackage{color}
\usepackage{amsmath,amssymb,epic,eepic,array}
\begin{document}
\renewcommand{\FirstPageOfPaper }{ 215}\renewcommand{\LastPageOfPaper }{ 220}

\title{Inner vacuum gap formation and drifting subpulses in pulsars}
\author{Janusz A. Gil \& George I. Melikidze}
\institute{Institute of Astronomy, University of Zielona G\'ora,
Lubuska 2, 65-265, Zielona G\'ora, Poland} \maketitle

\newcommand{\dmdt}{{\mbox{{\rm M}$_{\odot}$}} {\rm yr}$^{-1}$}
\newcommand{\gcc}{{\rm g} \, {\rm cm}^{-3}}
\newcommand{\rn}{$\rho_{\rm nuc}$}
\newcommand{\rd}{$\rho_{\rm drip}$}
\def\be{\begin{equation}}
\def\ee{\end{equation}}
\def\lesssim{\raisebox{-0.3ex}{\mbox{$\stackrel{<}{_\sim} \,$}}}
\def\gtrsim{\raisebox{-0.3ex}{\mbox{$\stackrel{>}{_\sim} \,$}}}

\begin{abstract}
The problem of formation of the inner vacuum gap in neutron stars
with ${\bf\Omega}\cdot{\bf B}<0$ is considered. It is argued by
means of the condition $T_i/T_s>1$, where $T_i$ is the melting
temperature of surface $^{56}_{26}$Fe ions and $T_s$ is the actual
temperature of the polar cap surface, that the inner vacuum gap
can form, provided that the actual surface magnetic field is
extremaly strong ($B_s\gtrsim 10^{13}$~G) and curved (${\cal
R}<10^6$~cm), irrespective of the value of dipolar component
measured from the pulsar spin down rate. The calculations are
carried out for pulsars with drifting subpulses and/or periodic
intensity modulations, in which the existence of the quasi steady
vacuum gap discharging via ${\bf E}\times{\bf B}$ drifting sparks
is almost unavoidable.
\end{abstract}

 \section{Introduction}

The subpulses in single pulses of a number of pulsars change phase
systematically between adjacent pulses, forming apparent
driftbands of the duration from several to few tenths of pulse
periods. The subpulse intensity is also systematically modulated
along driftbands, typically increasing towards the pulse centre.
In some pulsars, which can be associated with the central cut of
the line-of-sight trajectory, only periodic intensity modulation
is observed, without a systematic change of subpulse phase. On the
other hand, the clear subpulse driftbands are found in pulsars
with grazing line-of-sight trajectories. These characteristics
strongly suggest an interpretation of this phenomenon as a system
of subpulse associated beams rotating slowly around the magnetic
axis. Ruderman \& Sutherland (1975; hereafter RS75) proposed a
natural explanation of the origin of subpulse drift, that involved
a number of isolated ${\bf E}\times{\bf B}$ drifting sparks
discharging the quasi steady vacuum gap formed above the polar cap
of the so-called anti-pulsar with ${\bf\Omega}\cdot{\bf B}<0$, in
which $^{56}_{26}$Fe ions were strongly bound at the surface.
Although the original idea of RS75 associating the rotating
sub-beams with the circulating sparks is still regarded as the
best qualitative model of drifting subpulse phenomenon, their
vacuum gap was later demonstrated to suffer from the so-called
binding energy problem (for review see \cite{as91}). In fact, the
cohesive energies of $^{56}_{26}$Fe ions used by RS75 proved to be
largely overestimated and the inner vacuum gap envisioned by RS75
was impossible to form. However, it is worth to emphasize that
RS75 considered the canonical surface dipolar magnetic fields with
values determined from the pulsar spindown rate, although they
implicitly assumed small radii of curvature ${\cal R}\sim
10^6$~cm, inconsistent with purely dipolar field. Recently Gil \&
Mitra (2001; hereafter GM01) revisited the binding energy problem
an found that formation of vacuum gap is in principle possible,
although it requires extremely strong non dipolar surface magnetic
field $B_s=bB_d$, where the coefficient $b>>1$ in typical pulsar,
$B_d=6.4\times 10^{19}(P\dot{P})^{0.5}{\rm G}=2\times
10^{12}(P\dot{P}_{-15})^{0.5}$~G is the dipolar field at the pole,
$P$ is the pulsar period in seconds, $\dot{P}$ is the period
derivative and $\dot{P}_{-15}=\dot{P}/10^{-15}$.

In a superstrong surface magnetic field $B_s>0.1B_q$, where
$B_q=4.414\times 10^{13}$~G, the asymptotic approximation of
\cite{e66} used by RS75 in derivation of the height of quasi
steady vacuum gap is no longer valid. In fact, in such strong
field the high energy $E_f=\hbar\omega$ photons produce
electron-positron pairs at or near the kinematic threshold
$\hbar\omega=2mc^2/\sin\theta$, where $\sin\theta=h/{\cal R}$, $h$
is the gap height, and ${\cal R}={\cal R}_610^6$~cm is the radius
of curvature of surface magnetic field lines (e.g. \cite{dh83}),
$\hbar$ is the Planck constant, $c$ is the speed of light, $m$ and
$e$ are the electron mass and charge, respectively. The vacuum gap
formed under such conditions was called the Near Threshold Vacuum
Gap (hereafter NTVG) by GM01. They considered two kinds of high
energy seed photons dominating the $e^-e^+$ pair production: the
Curvature Radiation (CR) photons with energy
$\hbar\omega=(3/2)\hbar\gamma^3c/{\cal R}$ (RS75; \cite{zq96}),
and resonant Inverse Compton Scattering (ICS) photons with energy
$\hbar\omega=2\gamma\hbar eB_s/mc$ (\cite{zq96;zetal97}), where
$\gamma$ is a typical Lorentz factor of particles within the gap.
The corresponding vacuum gap is called the Curvature Radiation
dominated (CR-NTVG) and the Inverse Compton Scattering dominated
(ICS-NTVG), respectively. GM01 estimated the characteristic
heights of both CR-NTVG and ICS-NTVG. In this paper we will
further refine these estimates by including the general
relativistic (GR) effects of inertial frame dragging (IFD) and
considering the heat flow conditions within the thin uppermost
surface layer of the polar cap. Moreover, we will use broader
range of cohesive energies of surface $^{56}_{26}$Fe ions. The
obtained VG models are applied to pulsars with drifting subpulses
and/or periodic intensity modulations, in which the presence of
${\bf E}\times {\bf B}$ drifting spark discharges seems almost
unavoidable (Deshpande \& Rankin 1999; 2001; Vivekanand \& Joshi
1999 and Gil \& Sendyk 2000).

\section{Near threshold vacuum gap formation}

If the cohesive energy of $^{56}_{26}$Fe ions is large enough to
prevent them from thermionic (this section) or field emission
(Appendix A), a vacuum gap forms right above the polar cap.
\cite{jlk01} pointed out that the GR effect of IFD
(\cite{mt92;mh97}) should affect the RS75 type models with a
vacuum gap above the polar cap. Although \cite{jlk01} did not
investigate the problem, they implicitly suggested that the
electric fields distorted by GR effects make formation of
``starved'' inner magnetospheric regions even more difficult than
in the flat space case. However, Zhang et al. (2000) demonstrated
that although the GR-IFD effect is small, it nevertheless slightly
helps formation of VG above the polar caps. In other words, the GR
modified potential drop within the VG is slightly lower than in
the flat space case. Below we confirm this finding for NTVG
conditions, with a very strong and complicated surface magnetic
field.

The gap electric field ${\bf E}_\|$ along ${\bf B}_s$ results from
a deviation of the local charge density $\rho\approx 0$ from the
co-rotational charge density $\rho_c=(\zeta/\alpha)\rho_{GJ}$,
where $\rho_{GJ}=-{\bf\Omega}\cdot{\bf B}_s/(2\pi c)$ is the flat
space-time Goldreich-Julian (1969) charge density,
$\Omega=2\pi/P$, $\zeta=1-\kappa_g$,
$\kappa_g\sim(r_g/R)(I/MR^2)$, $\alpha=(1-r_g/R)^{1/2}$ is the
redshift factor, $r_g$ is the gravitational radius, $M$ is the
neutron star mass and $I$ is the neutron star moment of inertia.
The potential $V$ and electric fields $E_\|$ within a gap are
determined by GR-analog (\cite{mt92}) of the one dimensional
Poisson's equation $(1/\alpha)d^2V/dz^2=-4\pi(\rho-\rho_c)=2\Omega
B_s\zeta/(c\alpha)$, with a boundary condition $E_\|(z=h)=0$,
where $h$ is the height of infinitesimal gap. The solution of
Poisson's equation gives
\begin{equation}
E_{\|}(z)=\zeta(2 \Omega \frac {B_s} {c}) (h-z)
\end{equation}
and
\begin{equation}
\Delta V=\zeta\frac{\Omega B_s}{c}h^2 .
\end{equation}
In further calculations we will adopt a typical
value of the correction factor $\zeta\sim 0.85$ (corresponding to
$M=1M_\odot$, $R=10^6$~cm and $I=10^{45}{\rm g\ cm}^2$), although
its value can be as low as 0.73 (\cite{zhm00}). Thus, the
potential drop within the actual gap can be 15 to 27 percent lower
than in the conventional RS75 model.

The polar cap surface of the anti-pulsar (${\bf\Omega}\cdot{\bf
B}<0$) is heated by a back-flow of relativistic electrons
(accelerated in the parallel electric field $E_\|$) to the
temperature $T_s=k^{1/4}(e\Delta V\dot{N}/\sigma\pi r_p^2)^{1/4}$,
where $\Delta V$ is described by equation~(2), $\dot{N}=\pi
r_p^2B_s/eP$ is the Goldreich-Julian kinematic flux and the heat
flow coefficient $0.2\lesssim k<1$ is described in the Appendix B.
The thermal condition for the vacuum gap formation can be written
in the form $T_i/T_s>1$, where $T_s$ is the actual surface
temperature described above, and $T_i=\Delta\varepsilon_c/30$k is
the iron critical (melting) temperature above which $^{56}_{26}$Fe
ions are not bound on the surface (\cite{cr80;um95}), where
$\Delta\varepsilon_c$ is the cohesive energy of condensed
$^{56}_{26}$Fe matter in the neutron star surface and
k$=1.38\times 10^{-23}JK^{-1}$ is the Boltzman constant. The
properties of condensed matter in very strong magnetic fields
characteristic for the neutron star surface have been investigated
by many authors using different examination methods (for review
see \cite{um95}). There exists a lot of discrepancy in
determination of the cohesive energy $\Delta\varepsilon_c$ and in
this paper we refer to the two most useful papers, representing
the limiting extreme cases. Abrahams \& Shapiro (1991; AS91
henceforth) estimated $\Delta\varepsilon_c=0.91$~keV, 2.9 keV and
4.9 keV for $B_s=10^{12}$~G, $5\times 10^{13}$~G and $10^{13}$~G,
respectively. These values were approximated by \cite{um95} in the
form $\Delta\varepsilon_c\simeq(0.9{\rm keV})(B_s/10^{12}~{\rm
G})^{0.73}$, which leads to melting temperatures \begin{eqnarray}
T_i=(3.5\times 10^5{\rm K})(B_s/10^{12}{\rm G})^{0.73}=\nonumber \\
(6\times 10^5)b^{0.73}(P\dot{P}_{-15})^{0.36}~{\rm K}
.\end{eqnarray} On the other hand, Jones (1986; J86 henceforth)
obtained much lower cohesive energies
$\Delta\varepsilon_c=0.29$~keV, 0.60 keV and 0.92 keV for
$B_s=2\times 10^{12}$~G, $5\times 10^{12}$~G and $10^{13}$~G,
respectively. This can be approximated by
$\Delta\varepsilon_c\simeq(0.18\ {\rm keV})(B_s/10^{12}{\rm
G})^{0.7}$ and converted to melting temperatures  \begin{eqnarray}
T_i=(0.7\times 10^5{\rm K})(B_s/10^{12}{\rm G})^{0.7}=\nonumber \\
(1.2\times 10^5)b^{0.7}(P\cdot\dot{P}_{-15})^{0.36}
.\end{eqnarray} Below we consider the condition $T_i/T_s>1$, using
both expressions for melting temperatures described by
equations~(3) and (4), for CR- and ICS-dominated NTVG models,
separately.

\subsection{CR-NTVG}

In this case the gap height $h=h_{CR}$ is determined by the
condition that $h=l_{ph}$, where $l_{ph}\approx\sin\theta\ {\cal
R}=(B_\perp/B_s){\cal R}$ is the mean free path for pair
production by photon propagating at an angle $\theta$ to the local
surface magnetic field (RS75). The CR-NTVG model is described by
the following parameters: the height of quasi steady gap
\begin{equation}
h_{CR}=(3\times 10^3)\zeta^{-3/7}{\cal R}_6^{2/7}b^{-3/7}P^{3/14}
\dot{P}_{-15}^{-3/14}~{\rm cm} ,
\end{equation}
(notice typographical errors in eq.~[6] of GM01, which are
corrected here), the gap potential drop
\begin{equation}
\Delta V=(1.2\times 10^{12})\zeta^{1/7}{\cal R}_6^{4/7}b^{-1/7}P^{-1/14}
\dot{P}_{-15}^{1/14}\ {\rm V} ,
\end{equation}
and the surface temperature
\begin{equation}
T_s=(3.4\times 10^6)\zeta^{1/28}k^{1/4}{\cal R}_6^{1/7}b^{2/7}P^{-1/7}
\dot{P}_{-15}^{1/7} ~{\rm K}. 
\end{equation}
The thermal condition $T_i/T_s>1$ for the formation of CR-NTVG leads
to a family of critical lines on the $P-\dot{P}$ diagram (see
Fig.~1 in GM01) 
\begin{equation}
\dot{P}_{-15}\geq {\cal A}^2\zeta^{0.16}k^{1.14}{\cal R}_6^{0.64}b^{-2}P^{-2.3} ,
\end{equation}
where ${\cal A}=(2.7\times 10^3)^{1/2}=52$ for AS91 case (eq.~[3]) and
${\cal A}=(3.96\times 10^6)^{1/2}=1990$ for J86 case (eq.~[4]).
Alternatively, one can find a minimum required surface magnetic
field $B_s=bB_d$ expressed by the coefficient $b$ in the form 
\begin{equation}
b_{min}^{CR}={\cal A} \zeta ^{0.08}k^{0.57}{\cal R}_6^{0.32}P^{-1.15}
\dot{P}_{-15}^{-0.5} .
\end{equation}

\subsection{ICS-NTVG}

In this case the gap height $h=h_{ICS}$ is determined by the
condition $h=l_{ph}\sim l_e$, where $l_e$ is the mean free path of
the electron to emit a photon with energy
$\hbar\omega=2\gamma\hbar eB_s/mc$ (GM01, \cite{zhm00}). The
ICS-NTVG model is described by the following parameters: the
height of quasi steady gap 
\begin{equation}
h_{ICS}=(5\times 10^3)\zeta^{0.14}k^{-0.07}{\cal
R}_6^{0.57}b^{-1}P^{-0.36}\dot{P}_{-15}^{-0.5}~{\rm cm},
\end{equation}
the gap potential drop 
\begin{equation}
\Delta V=(5.2\times 10^{12})\zeta^{0.72}k^{-0.14}{\cal
R}_6^{1.14}b^{-1}P^{-1.22}\dot{P}_{-15}^{-0.5}~{\rm V},
\end{equation}
and the surface temperature 
\begin{equation}
T_s=(4\times 10^6)\zeta^{0.18}k^{0.25}{\cal R}_6^{0.28}P^{-0.43} ~{\rm K}. 
\end{equation}
The thermal condition $T_i/T_s>1$ for the formation of ICS-NTVG
leads to a family of critical lines on the $P-\dot{P}$ diagram
(see Fig.~1 in GM01) 
\begin{equation}
\dot{P}_{-15}\geq {\cal B}^2\zeta^{0.5}k^{0.7}{\cal R}_6^{0.8}b^{-2}P^{-2.2} ,
\end{equation}
where ${\cal B}=(2\times 10^2)^{1/2}=14$ for AS91 case (eq.~[3]) and
${\cal B}=(1.69\times 10^4)^{1/2}=130$ for J86 case (eq.~[4]).
Alternatively, one can find a minimum required surface magnetic
field $B_s=bB_d$ expressed by the coefficient $b$ in the form 
\begin{equation}
b_{min}^{ICS}={\cal B}\zeta^{0.25}k^{0.34}{\cal
R}_6^{0.39}P^{-1.1}\dot{P}_{-15}^{-0.5} .
\end{equation}

\subsection{NTVG development in pulsars with drifting subpulses}

GM01 examined the $P-\dot{P}$ diagram (their Fig.~1) with 538
pulsars from the Pulsar Catalog (\cite{tetal93}) with respect to
the possibility of VG formation by means of the condition
$T_i/T_s$, with $T_i$ determined according to AS91 (eq.~[3]). They
concluded that formation of VG is in principle possible, although
it requires very strong and curved surface magnetic field
$B_s=b\cdot B_d\gtrsim 10^{13}$~G, irrespective of the value of
$B_d$. Here we reexamine this problem by means of equations~(9)
and (14), with melting temperatures corresponding to both the AS91
case (${\cal A}=52$ and ${\cal B}=14$) and J86 case (${\cal
A}=1990$ and ${\cal B}=130$), for CR and ICS seed photons,
respectively. Therefore, we cover practically almost the whole
range of melting temperatures between the two limiting cases,
which differ by a factor of five between each other. We adopt the
GR-IFD correction factor $\zeta=0.85$ (\cite{hm98;zh00}), the
normalized radius of curvature of surface magnetic field lines
$0.01\leq{\cal R}_6\leq 1.0$ (GM01 and references therein) and the
heat flow coefficient $0.2\leq k\leq 1.0$ (Appendix B).

The results of calculations of NTVG models for 42 pulsars with
drifting subpulses and/or periodic intensity modulations (after
Rankin 1986) are presented in Fig.~1. We calculated the ratio
$B_s/B_q=0.0453\ b(P\dot{P}_{-15})^{0.5}$, where $B_q=4.414\times
10^{13}$~G and the coefficient $b$ is determined by equation~9 and
equation~14 for CR and ICS seed photons, respectively. Noting that
the functional dependence on $P$ and $\dot{P}_{-15}$ in both
equations is almost identical, we sorted the pulsar names (shown
on the horizontal axes) according to the increasing value of the
ratio $B_s/B_q$ (shown on the vertical axes). Calculations were
carried out for $B_s/B_q\geq 0.1$, since below this value the NT
conditions are no longer valid. On the other hand, the values of
$B_s/B_q$ are limited from above by the photon splitting
threshold, which is roughly about $10^{14}$~G (see also
astro-ph/0102097, \cite{bh98;bh01;z01}). Therefore, the physically
plausible NTVG models lie within the grey areas, which correspond
to surface magnetic fields $4.4\times 10^{12}~{\rm G}\leq B_s\leq
10^{14}$~G. The left hand side of Fig.~1 corresponds to AS91 case
(${\cal A}=52$ and ${\cal B}=14$ in eqs.~[9] and [14],
respectively) and the right hand side of Fig.~1 corresponds to J86
case (${\cal A}=1930$ and ${\cal B}=130$ in eqs.~[9] and [14],
respectively). Four panels in each side of the figure correspond
to different values of the radius of curvature ${\cal R}_6=1.0$,
0.1. 0.05 and 0.01 from top to bottom, respectively (indicated in
left upper corner of each panel). Two sets of curved lines in each
panel correspond to the CR (thin upper lines) and ICS (thick lower
lines) seed photons, respectively. Three different lines within
each set correspond to different values of the heat flow
coefficient $k=1.0$ (dotted), $k=0.6$ (dashed) and $k=0.2$ (long
dashed).

\begin{figure*}
\centerline{\psfig{file=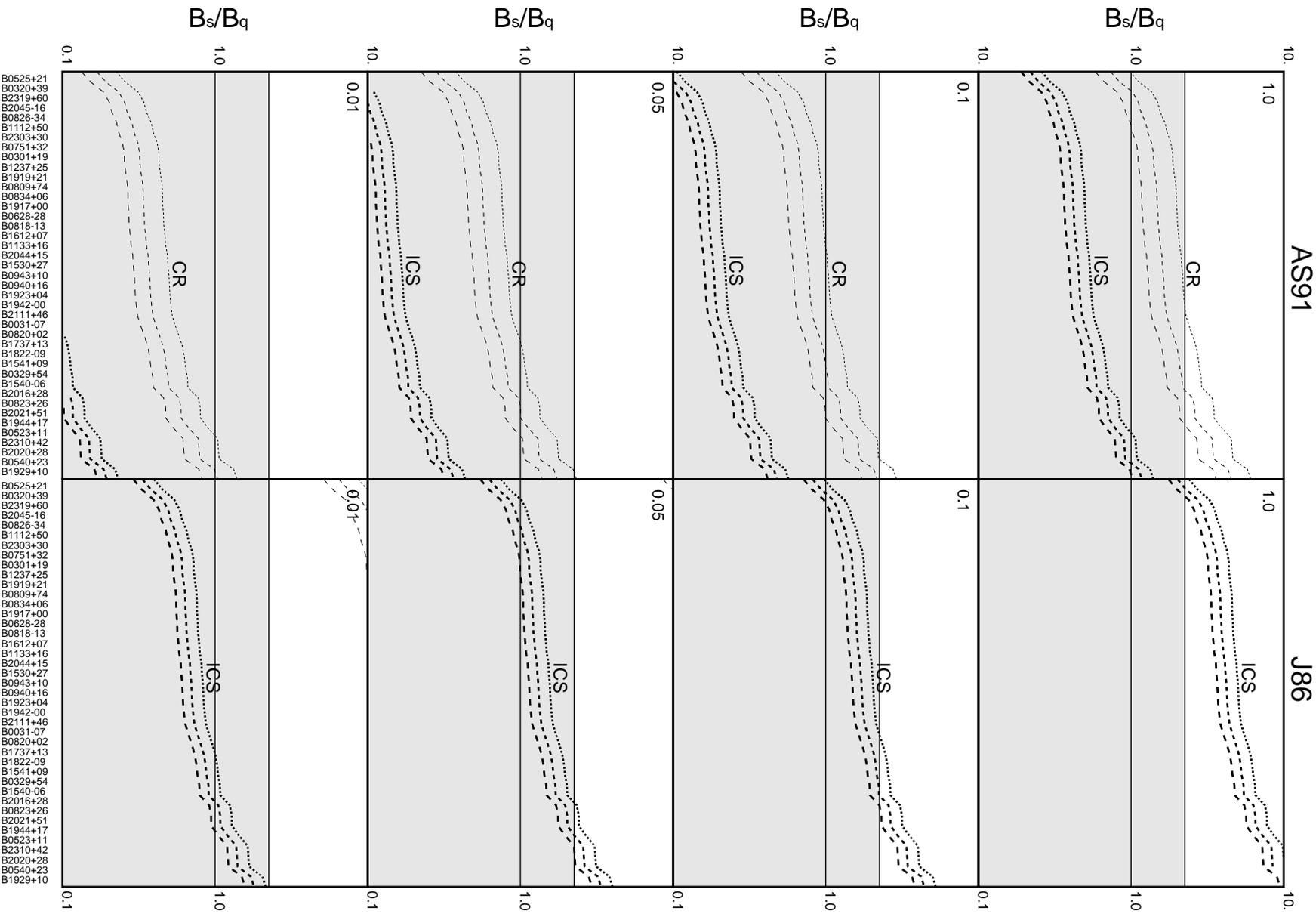,width=16.cm,clip=} } 
\caption{Models of NTVG for 42 pulsar with drifting subpulses and/or periodic
intensity modulations (see text for explanations). \label{image}}
\end{figure*}

A visual inspection of the model curves within grey areas in
Fig.~1 shows that in AS91 case the ICS-NTVG is favored for larger
radii of curvature ${\cal R}_6>0.05$, while the CR-NTVG requires
lower values of ${\cal R}_6<0.1$. In J86 case, only the ICS-NTVG
corresponding to ${\cal R}_6<0.1$ can develop. If the actual
cohesive energies correspond to some intermediate case between the
AS91 and J86 cases, they will also be associated with the
ICS-NTVG. Therefore, we can generally conclude that the ICS-NTVG
model is apparently favored in pulsars with drifting subpulses.
The CR-NTVG is also possible although it requires an extremely
strong ($B_s/B_q\sim 1$) and/or extremely curved (${\cal R}_6\sim
0.01$) surface magnetic field at the polar cap.

\section{Conclusions}

There is a growing evidence that the radio emission of pulsars
with systematically drifting subpulses (grazing cuts of the
line-of-sight) or periodic intensity modulations (central cuts of
the line-of-sight) is based on the inner vacuum gap developed just
above the polar cap (\cite{dr99;dr01;vj99;gs00}). To overcome the
binding energy problem (\cite{xqz99;xzq01}) put forward an
attractive but exotic conjecture that pulsars showing the drifting
subpulses represent bare polar cap strange stars (BPCSS) rather
than neutron stars. However, as demonstrated in this paper,
invoking the BPCSS conjecture is not necessary to explain drifting
subpulse phenomenon. The quasi steady vacuum gap, with either
curvature radiation or inverse Compton scattering seed photons,
can easily form in pulsars with ${\bf\Omega}\cdot{\bf B}<0$,
provided that the actual surface magnetic field at the polar cap
is extremely strong $B_s\sim 10^{13}$~G and curved ${\cal
R}<10^6$~cm, irrespective of the value of dipolar component
measured from the pulsar spindown rate.

The pulsars with drifting subpulses and/or periodic intensity
modulation do not seem to occupy any particular region of the
$P-\dot{P}$ diagram (see Fig.~1 in GM01). Rather, they are spread
uniformly all over the $P-\dot{P}$ space, at least for typical
pulsars. Therefore, it seems tempting to propose that radio
emission of all pulsars is driven by vacuum gap activities. An
attractive property of such proposition is that the nonstationary
sparking discharges induce the two-steam instabilities that
develop at relatively low altitudes (\cite{u87;am98}), where the
pulsars radio emission is expected to originate
(\cite{c78;kg97;kg98}). It is generally believed that the high
frequency plasma waves generated by the two-stream instabilities
can be converted into coherent electromagnetic radiation at pulsar
radio wavelengths (e.g. \cite{mgp00}). In such a scenario, all
radio pulsars would require a strong, non-dipolar surface magnetic
field at their polar caps (e.g. Gil et al. 2002 a, b).

\begin{acknowledgements}
We gratefully acknowledge the support by the Heraeus foundation.
We are grateful to Prof. V. Usov and dr B. Zhang for helpful
discussions. We also thank E. Gil, U. Maciejewska, and M. Sendyk
for technical help.
\end{acknowledgements}

\begin{appendix}

\section{Field emission}

The vacuum gap can form in pulsars with ${\bf\Omega}\cdot{\bf
B}<0$ if the actual surface temperature $T_s$ is not high enough
to liberate $^{56}_{26}$Fe ions from the polar cap surface by
means of thermal emission. Now we will examine the field (cold
cathode) emission, which is possibly important when thermionic
emission is negligible (e.g. \cite{zh00}). The maximum electric
field (along ${\bf B}_s$) at the NS surface
$E_\|(max)=\zeta(4\pi/cP)B_sh_{ICS}=(1.25\times
10^9)\zeta^{0.86}b^{-1}{\cal R}_6^{0.57}P^{0.14}~{\rm V}/{\rm
cm}$, or taking into account that $\zeta=0.85$ and $b\gtrsim
2(P\cdot\dot{P}_{-15})^{-0.5}$, $E_\|(max)\leq(5\times 10^8){\cal
R}_6^{0.57}P^{0.64}\dot{P}_{-15}^{0.5}$~V/cm. The critical
electric field needed to pull $^{56}_{26}Fe$ iron ions from the NS
surface is $E_\|(crit)=(8\times 10^{12})(\Delta\epsilon/26~{\rm
keV})^{3/2}$~V/cm (\cite{as91;um95;um96}), where the cohesive
(binding) energy of iron ions in strong surface magnetic field
$B_s\sim 10^{13}$~G is $\Delta\epsilon=4.85$~keV (\cite{as91}).
Thus, $E_\|(crit)=6.4\times 10^{11}~{\rm V/cm}>>E_\|\sim
10^9$~V/cm and no field emission occurs. It is possible that the
cohesive energy is largely overestimated and $\Delta\varepsilon_c$
can be much smaller than about 4 keV even at $B_s>10^{13}$~G. We
can thus ask about the minimum value of $\Delta\varepsilon_c$ at
which the field emission is still negligible. By direct comparison
of $E_\|(max)$ and $E_\|(crit)$ we obtain
$\Delta\varepsilon_c>40x^{0.67}$~eV, where $x={\cal
R}_6^{0.57}P^{0.64}\dot{P}_{-15}^{0.5}$ is of the order of unity.
Therefore, contrary to the conclusion of \cite{jlk01}, no field
emission is expected under any circumstances.

\section{Heat flow conditions at the polar cap surface}

Let us consider the heat flow conditions within the uppermost
surface layer of the pulsar polar cap above which NTVG can
develope. The basic heat flow equation is (e.g. \cite{ec89})
\begin{equation}
C\frac{\partial T}{\partial t}=\frac{\partial}{\partial
x}\left(\kappa_\|\frac{\partial T}{\partial x}\right) ,
\end{equation}
where $C$ is the heat capacity (per unit volume) and
$\kappa_\|>>\kappa_\perp\sim 0$ is the thermal conductivity along
($\|$) surface magnetic field lines (which are assumed to be
perpendicular ($\perp$) to the polar cap surface). The heating of
the surface layer of thickness $\Delta x$ is due to sparking
avalanche with a characteristic development time scale $\Delta t$.
We can write approximately $\partial T/\partial t\approx T/\Delta
t$, $\partial T/\partial x\approx T/\Delta x$, $\partial/\partial
x(\partial T/\partial x)\approx T/\Delta x^2$ and thus $C/\Delta
t\approx \kappa_\|\Delta x^2$. Therefore, the crust thickness
$\Delta x$ heated during $\Delta t$ is approximately $\Delta x
\approx\left({\kappa_\|}\Delta t/{C}\right)^{1/2}$ (within an
uncertainty factor $\delta x$ of 2 or so), where the time scale
corresponds to the spark development time $\Delta t\approx
10\mu{\rm s}$ (RS75). The energy balance equation is
$Q_{heat}=Q_{rad}+Q_{cond}$, where $Q_{heat}=en_{GJ}\Delta
V_{max}$ (e.g. RS75), $Q_{rad}=\sigma T_s^4$ ($T_s$ is the actual
surface temperature and $\sigma=5.67\times 10^{-5}~{\rm erg\
cm}^{-2}K^{-4}s^{-1}$), and $Q_{cond}=-\kappa_\|\partial
T/\partial x\approx -\kappa_\|T_s/\Delta x$. We can now define the
heat flow coefficient $k=Q_{rad}/Q_{heat}<1$, which describes
deviations from the black-body conditions on the surface of the
sparking polar cap. In other words, the value of $k$ describes the
amount of heat conducted beneath the polar cap which cannot be
transferred back to the surface from the penetration depth $\Delta
x$ during the time-scale $\Delta t$. The coefficient $k$ can be
written in the form
 \begin{equation}
k=\frac{1}{1+\kappa_\|/(\sigma T_s^3\Delta
x)}=\frac{1}{1+(\kappa_\|C)^{1/2}/(\sigma T_s^3\Delta t^{1/2})} ,
\end{equation}
whose value can be estimated once the parameters $C$,
$\kappa_\|$ and $T_s$ as well as $\Delta x$ or $\Delta t$ are
known.

\begin{figure}
\centerline{\psfig{file=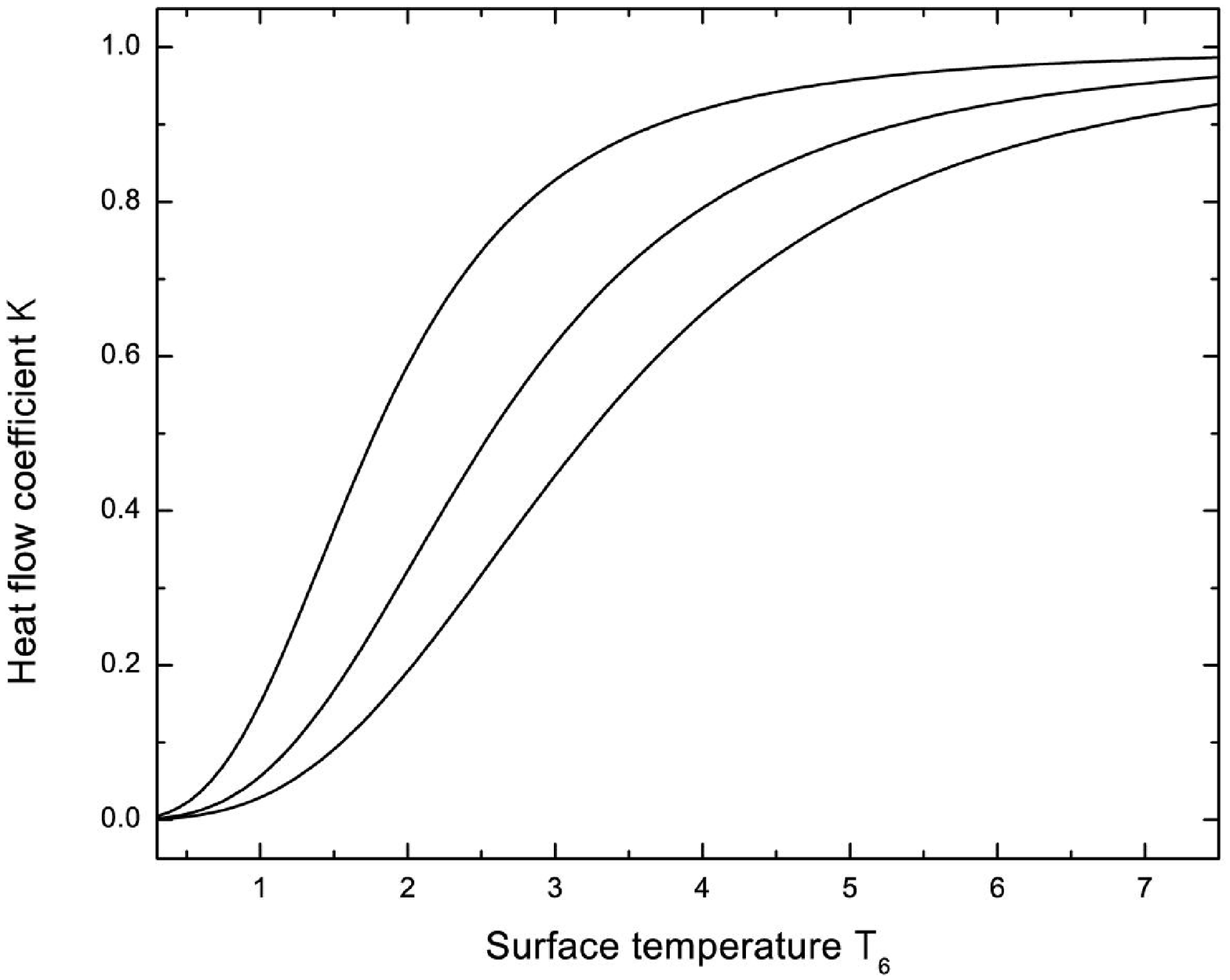,width=8.8cm,clip=} } 
\caption{Heat flow coefficient $k$ versus surface temperature $T_6$ (see text
for explanations). \label{image}}
\end{figure}

The matter density at the neutron star surface composed mainly of
$^{56}_{26}$Fe ions (e.g. \cite{um95}) is $\rho(B_s)\simeq 4\times
10^3(B_s/10^{12}~{\rm G})^{6/5}{\rm g\ cm}^{-3}$ (\cite{fetal77}).
Thus for $B_s\sim (1\div 3)\times 10^{13}$~G we have $\rho\sim
(0.6\div 2.4)\times 10^5{\rm g\ cm}^{-3}$. The thermal energy
density of $^{56}_{26}$Fe is $U_{Fe}\simeq 2.2\times
10^{19}\rho_5T_8\ {\rm erg\ cm}^{-3}$ (\cite{ec89}), where
$\rho_5=\rho/(10^5 {\rm g\ cm}^{-3})$ and $T_8=T_s/(10^8~{\rm
K}$). Thus, the heat capacity
$C=dU_{Fe}/dt=10^{-8}dE_{Fe}/dT_8=(1\div 5)\times 10^{11}\rho_5\
{\rm erg\ cm}^{-3} {\rm K}^{-1}$. For $\rho_5\approx 1$ we have
$C\approx 2\times 10^{11}{\rm erg\ cm}^{-3}{\rm K}^{-1}$. The
longitudinal thermal conductivity can be estimated as
$\kappa_\|\approx (2\div 4)\times 10^{13}\ {\rm erg\ s}^{-1}{\rm
cm}^{-1}{\rm K}^{-1}$ (see Fig.~5 in \cite{p99}). Thus
$\kappa_\|/C\approx 1.5\times 10^2\ {\rm cm}^2$ and the
penetration depth $\Delta x\approx 10\Delta t^{1/2}s^{-1/2}$~cm.
Since $\Delta t\sim 10 \mu =10^{-5}$~s (spark characteristic time
scale) then $\Delta x\approx 0.03$ ~cm. More generally, the
penetration depth can be written as 
\begin{equation}
\Delta x=0.03\,\delta
x\left(\frac{\kappa_{13}}{C_{11}}\right)^{1/2} ,
\end{equation}
where $\kappa_{13}=\kappa_\|/10^{13}\approx 2\div 4$ and
$C_{11}=C/10^{11}\approx 1\div 5$ and the uncertainty factor
$\delta x\approx 0.5\div 2$. Now the heat flow coefficient can be
written as 
\begin{equation}
k=\frac{1}{1+5.6\Delta/T_6^3} ,\end{equation} where
$T_6=T_s/10^6$ and $\Delta=(\kappa_{13}C_{11})^{1/2}/\delta
x\approx 0.7\div 6.3$. Figure 2 shows variations of the heat flow
coefficient $k$ versus the surface temperature $T_6$ (in $10^6$~K)
for three values of $\Delta=1$ (upper curve), $\Delta=3$ (middle
curve) and $\Delta=6$ (lower curve). As one can see, for realistic
surface temperatures of few times $10^6$~K, the values of the heat
flow coefficient $k$ are in the range $0.2\div 0.9$. \cite
{gmm02a} found from independent considerations that in PSR
B0943+10 the heat flow coefficient $k<0.8$, in consistency with
the results obtained in this paper.
\end{appendix}


\clearpage

\end{document}